\documentclass[proceedings, preprint]{rmaa}

\usepackage{psfrag,color}

\usepackage{hyperref}

\SetYear{2020}
\SetConfTitle{Robotic Observatories Workshop}

\title{The Zadko Observatory}

  \author{J. A. Moore,\altaffilmark{1,2} 
					B. Gendre, \altaffilmark{1,2} 
					D.M. Coward, \altaffilmark{1,2} 
					H. Crisp, \altaffilmark{1,2} 
					and A. Klotz \altaffilmark{3,2}
					}

\altaffiltext{1}{Department of Physics, University of Western Australia, 35 Stirling Highway, Crawley 6009, Australia
    (john.moore@uwa.edu.au).}
\altaffiltext{2}{Australian Research Council Centre of Excellence, OzGrav}
\altaffiltext{3}{IRAP, 14 avenue Edouard Belin, 31400 Toulouse, France.}

\shortauthor{Moore et al.}
\shorttitle{Zadko Observatory}

\listofauthors{J. A. Moore, B. Gendre , D.M. Coward, H. Crisp \& A. Klotz}

\indexauthor{Moore, J. A.}
\indexauthor{Gendre, B.}
\indexauthor{Coward, D.M.}
\indexauthor{Crisp, H}
\indexauthor{Klotz, A.}

\abstract{The 1.0 metre f/4 fast-slew Zadko Telescope was installed in June 2008 approximately seventy  kilometres north of Perth at Yeal, in the Shire of Gingin, Western Australia. Since the Zadko Telescope has been in operation it has proven its worth by detecting numerous Gamma Ray Burst afterglows, two of these being the most distant `optical transients' imaged by an Australian telescope. Other projects include a contract with the European Space Agency (ESA) to image potentially hazardous near Earth asteroids (2019), monitoring space weather on nearby stars (2019), and photometry of a transit of Saturn's moon Titan (2018). Another active Zadko Telescope project is tracking Geo-stationary satellites and attempting to use photometry to classify various space debris (defunct satellites). The Zadko Telescope's importance as a potential tool for education, training, and public outreach cannot be underestimated, as the global awareness of the importance of astronomy (and space science) as a context for teaching science continues to increase. An example of this was the national media coverage of its contribution to the discovery of colliding neutron stars in 2017, capturing the imagination of the public. In this proceeding, I will focus on the practical aspects of managing a robotic Observatory, focusing on the sustainability of the Observatory and the technical management involved in hosting different commercial projects. I will review the evolution of the Observatory, from its early, single instrument, state to its current multi-telescope and multi-instrument capabilities. I will finish by outlining the future of the Observatory and the site.}

\resumen{El telescopio Zadko de 1,0 metros f/4 y de r\'apida capacidad de apuntado se instal\'o en junio de 2008 a unos 70 km al N de Perth en Yeal, en la Comarca de Gingin (Australia Occidental). Desde que el telescopio Zadko ha estado en funcionamiento, ha detectado numerosas contrapartidas \'opticas de estallidos de rayos-gamma (GRBs), dos de los cuales se cuentan entre las fuentes m\'as lejanas registradas por un telescopio australiano. Otros proyectos incluyen un contrato con la Agencia Espacial Europea para la adquisici\'on de im\'agenes de asteroides potencialmente peligrosos cercanos a la Tierra (2019), habi\'endose tambi\'en estudiado las estrellas m\'as cercanas (2019), as\'i como realiz\'andose fotometr\'ia de un tr\'ansito de la luna Titán de Saturno (2018). Otro proyecto activo del telescopio es estar rastreando sat\'elites geoestacionarios e intentando utilizar la fotometr\'ia para clasificar basura espacial. El telescopio Zadko es igualmente importante para la educaci\'on y formaci\'on, en el contexto de divulgaci\'on de la ciencia, ya que la importancia de la astronom\'ia (y ciencias del espacio) contin\'ua aumentando. Un ejemplo de esto fue la amplia cobertura en los medios dada a la detección de ondas gravitacionales producidas por una colisi\'on de estrellas de neutrones en 2017. En este trabajo nos centraremos en los aspectos pr\'acticos de la gesti\'on de un observatorio rob\'otico, haciendo \'enfasis en la sostenibilidad del observatorio y la gesti\'on t\'ecnica involucrada para satisfacer los diferentes proyectos en los que se participa. 
Revisaremos la evoluci\'on del observatorio, desde su estado inicial e instalaci\'on de un \'unico instrumento hasta su estado actual con la capadidad de albergar m\'ultiples telescopios y su correspondiente instrumentaci\'on, para concluir describiendo los proyectos futuros del observatorio.}

\addkeyword{Telescopes}
\addkeyword{instrumentation: miscellaneous}

\begin{document}
\maketitle

\section{Introduction}
\begin{figure}[!b]\centering
\vspace{0pt}
  \includegraphics[width=7.5cm]{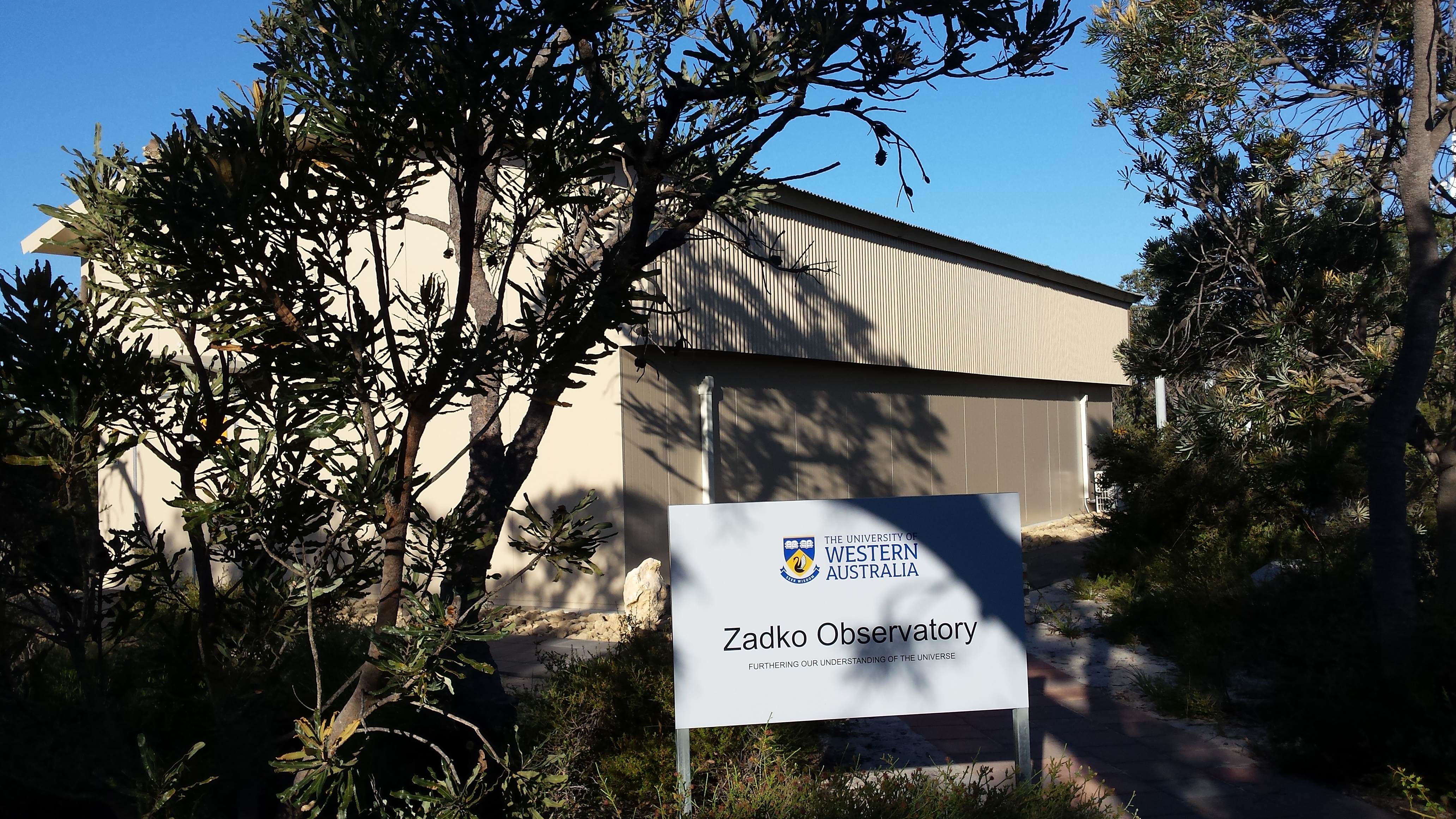}
  \caption{External view of the Zadko Observatory.}
  \label{fig:observatory}
\end{figure}

The Zadko Telescope \citep{cow10} is a 1.0 metre f/4 Ritchey-Chrétien telescope located in Western Australia, in the shire of Gingin (see Figure \ref{fig:observatory}). Constructed by DFM Engineering Inc. in Longmont Colorado USA and donated to the University of Western Australia by Resource Company Claire Energy CEO, Jim Zadko. The telescope has been in operation since 1st April 2009. Specifications of the Zadko Observatory are listed in Table \ref{table1}.

\section{Evolution of the Observatory}
\label{sec:intro}

\begin{figure*}[!t]\centering
\vspace{0pt}
  \includegraphics[scale=0.4]{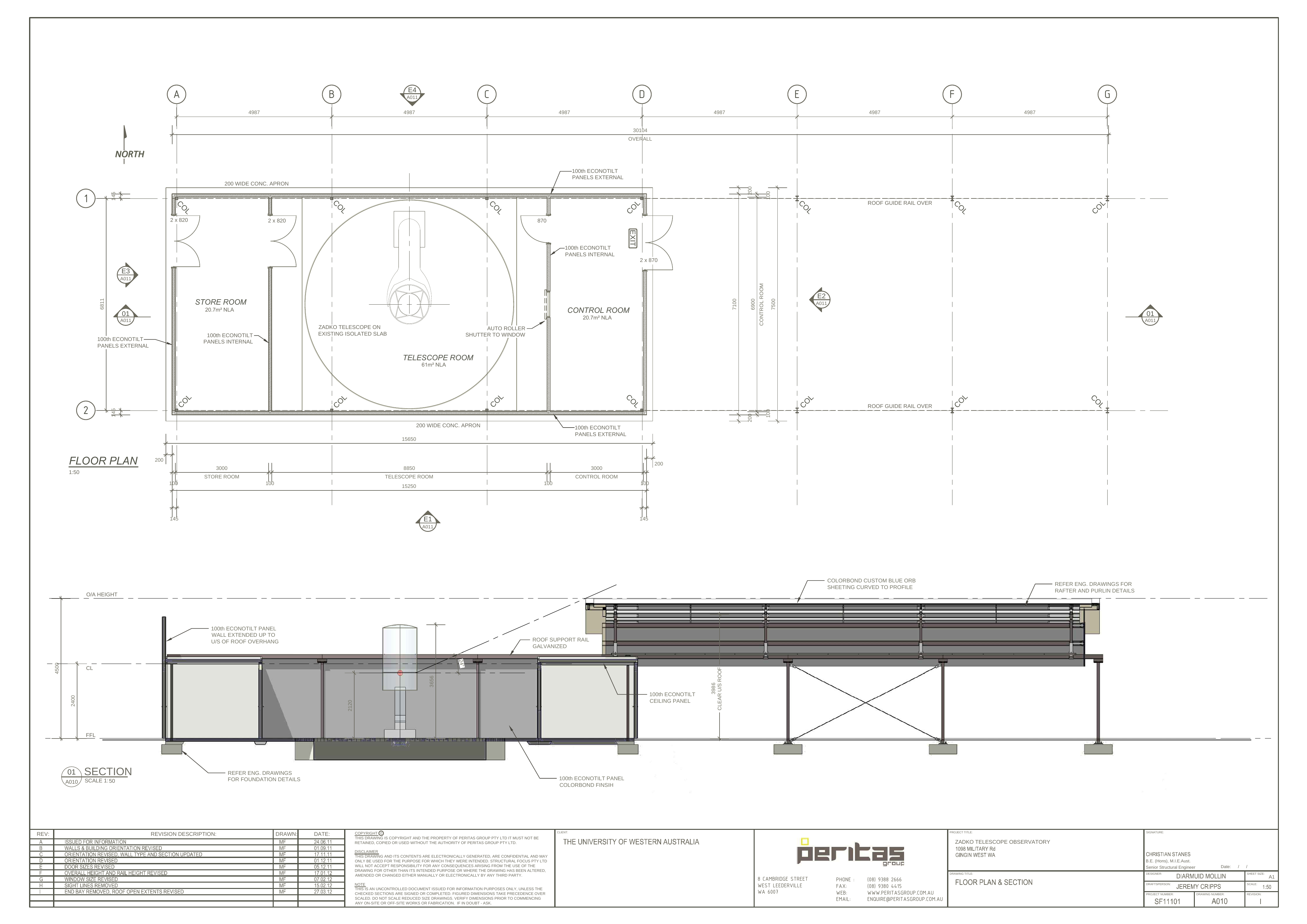}
  \hspace*{\columnsep}
  \caption{Zadko Observatory mechanical drawings, showing floor plans for top and side views, including the position of the 1-m Zadko Telescope.}
  \label{fig:widefig1}
\end{figure*}

Due to operational issues the original 6.7 metre fibreglass domed observatory which the Zadko Telescope \citep{cow10,cow17} was originally housed was replaced in 2011 with a state of the art purpose built robotic rolling roof autonomous observatory with a dedicated 21 m$^2$ constant temperature climate controlled operations room. In addition to this a climate controlled telescope service room to mirror the operation room was also added. This had the benefit of allowing for possible future conversion into a second control room when other telescopes were added to the increased 63 m$^2$ telescope viewing area. Robotic control of the rolling roof is performed using a PLC-Burgess system designed and manufactured by the electronics workshop at Observatoire de Haute-Provence. A brief system overview is explained here: The PLC is connected to the PC-Zosma via an Ethernet connection. The PLC receives commands from the PC-Zosma and sends status to the PC-Zosma through a socket connection. The weather station and the cloud sensor are connected to the serial ports of PC-Zosma. The PC processes this data and sends commands to the PLC in order to control the roof in a safe condition. All the safety devices are controlled by the PLC. In the event of a communication failure with the PC, the PLC closes the roof. Refer to Figure \ref{fig:widefig1} for floor plan of the Zadko Observatory.

\begin{table}[!bht]\centering
  \setlength{\tabnotewidth}{\columnwidth}
  \tablecols{2}
  \caption{Zadko Observatory physical parameters. } \label{table1}
  \begin{tabular}{lr}
    \toprule
    Property & Value \\
    \midrule
		Observatory code    & D20 \\
		Longitude           & 115$^o$ 42' 49'' E \\
		Latitude            &  31$^o$ 21' 24'' S \\
		Altitude            & 50 meters \\
		Seeing              & Variable (1-4\arcsec) \\
		magnitude  & R$\sim$21.5 (180s exposure) \\
		Zadko Camera              & QHY 160M \\
		Zadko Filters             & Sloan g', r', i', Clear \\
		Commercial clients  &   4 \\
    \bottomrule
  \end{tabular}
\end{table}

\subsection{Weather station}

A critical component of all robotic observatories is the ability to monitor site conditions using a variety of instruments. The Zadko Observatories instruments feed back to the PLC-Burgess and the TCS to signal when it is safe to open the roof and commence viewing. Alternately if a detrimental weather event is detected the roof will automatically close thereby protecting the telescopes from water, dust or smoke ingress.

The Saia Burgess Control PCD3 module that controls the roof enables bypass of any supporting computer allowing an immediate response to close the roof. The PCD3 module also acts as a failsafe in the event that any detector reports incoherent values or is lost the observatory will go into safe mode by closing the roof and forbidding it to re-open without human intervention.

Instruments that form part of the Zadko Observatories control are:

\begin{itemize}

\item A {\bf Hydreon RG-11 Rain Gauge}\footnote{https://rainsensors.com/} senses water hitting its outside surface using beams of infrared light. It uses the same sensing principle used in millions of automotive rain sensing windshield wiper controls. The RG-11 is optical - not mechanical, chemical, or conductive. Consequently, it is far more rugged, sensitive and reliable than any other technology. The sensor is extremely sensitive, and virtually immune to false trips. Yet, it is completely unaffected by jostling and motion. There are no exposed conductors to corrode, and no openings for bugs to crawl into. There is no place for leaves or other debris to collect. It is also inexpensive compared to other similar instruments on the market. The RG-11 detector has been in operation for over twelve months without error.

\item A {\bf Diffraction Limited Boltwood II Cloud Sensor}\footnote{http://diffractionlimited.com/product/boltwood-cloud-sensor-ii/} that measures the amount of cloud cover by comparing the temperature of the sky to the ambient ground level temperature. The sky temperature is determined by measuring the amount of radiation in the 8 to 14 micron infrared band. A large difference indicates clear skies, whereas a small difference indicates dense, low-level clouds. This allows the sensor to continuously monitor the {\bf clarity} of the skies, and to trigger appropriate alerts on your computer. The device also includes a {\bf moisture sensor} which detects rain and snow. To prevent false alarms due to frost or dew, a heater keeps the sensor slightly above ambient temperature. When rain or snow falls, the sensor is automatically heated to 70 degrees Celsius. This clears the sensor quickly when the precipitation ends, and ensures that the sky-measuring thermophile has a clear view. The Boltwood Cloud Sensor II measures {\bf wind speed}, using a specially-designed anemometer with no moving parts. The sensor will warn you when winds speeds are too high for safe operation of the observatory. Plus the sensor detects {\bf daylight} and can be set to automatically close the roof to prevent any possibility of sunlight entering the telescope. The sensor also measures {\bf humidity}, and provides a continuous readout of temperature, humidity, and dew point.

\item  A Vaisala weather control unit\footnote{http://www.vaisala.com} providing humidity, temperature, and wind data.

\item  A GPS antenna for precision (microsecond) timing

\item An external webcam allowing a clear view of the observatories roof status and sky to the horizon for remote manual inspection. 
 
\end{itemize}

All of these instruments are located on a six metre high mast external to the observatory at approximately three metre distance from the control room. A flashing amber light is installed at the top to signal roof movement as a safety measure.


\begin{figure}[!h!]\centering
  \includegraphics[width=7.5cm]{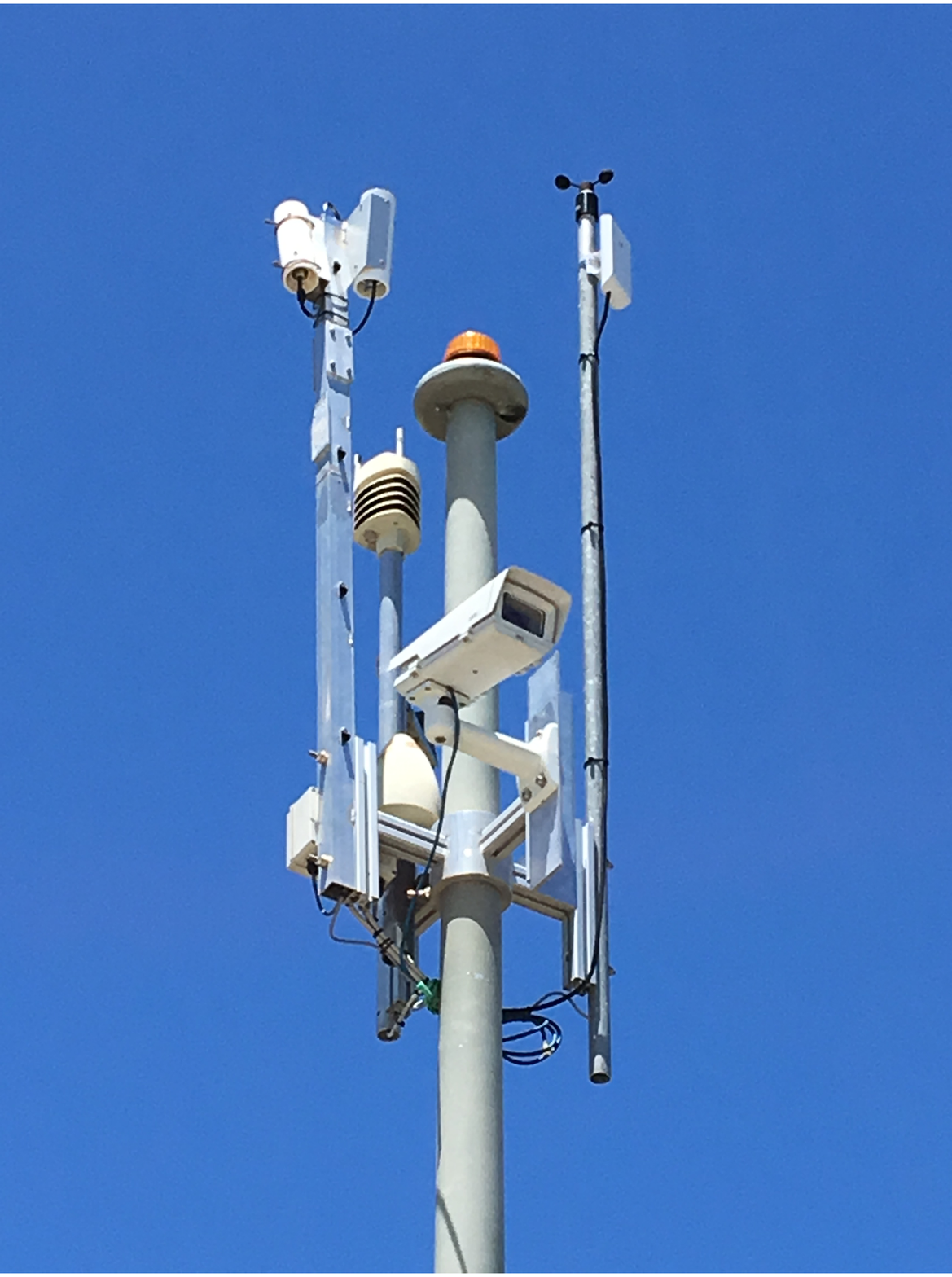}
  \caption{Mast housing weather station sensors for cloud, rain, temperature, humidity, wind speed/direction, internet camera and GPS antenna.}
  \label{fig:weather}
\end{figure}


\subsection{Mirror re-coating}

Due to the Zadko telescope important science capability it is essential that the primary and secondary astronomical reflecting surfaces (mirrors) are kept in an optimal state of cleanliness to reduce contaminants which can significantly degrade their reflectivity, IR emissivity, and light-scattering properties. In October 2018 the Zadko telescopes Primary and Secondary mirrors were removed and returned to Evaporated Metal Films Corporation in Ithaca NY, USA for specialized treatment including coating with standard enhanced aluminium (Al-SiOx/TiOx) with R average $\ge 93 \%$ over range 450 to 650 nm. 

This was an improvement on the mirrors original average reflectivity, which had an R average of 86-88\% or better with standard protected aluminium coating. 

As the Zadko Observatory is located in a hostile environment (high pollen and dust) at 50 metres above sea level surrounded by native bush and only 18 kilometres from the coast, the protective aluminium coating has served its purpose well and provided a ten-year useable life before re-coating became necessary. Due to the work required for removal and reinstallation of the mirrors plus the cost involved in the re-coating process, obtaining twice the recommended period for re-coating has been considered a bonus.

Further improvements to the telescope room environment will be investigated over the coming months to improve air quality and enhance the temperature control.

\begin{figure}[!t]\centering
  \includegraphics[width=7.5cm]{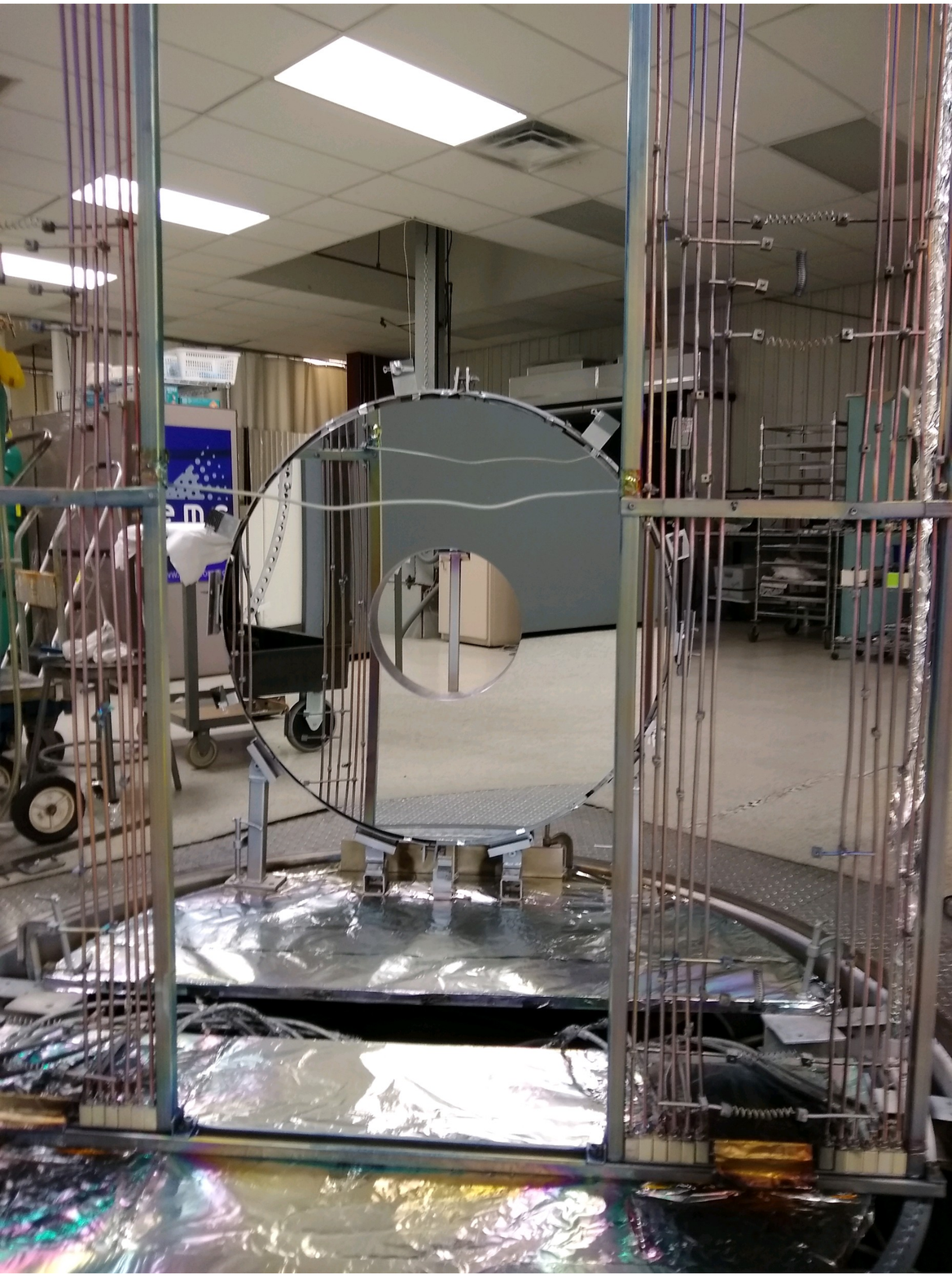}
  \caption{Zadko Telescope primary mirror after re-coating.}
  \label{fig:mirror}
\end{figure}

\section{Observatory management}
To maintain a fully autonomous observatory requires careful strategic planning including long-term goal setting and financial support plus a dedicated and committed team. Over the past decade the Zadko Observatory has steadily extended its research capability and has become part of several large collaborations, with an emphasis on transient astronomy and gravitational wave follow-up \citep{ant20,and17,abb17}. 

\subsection{Commercial clients}

With a view to future expansion the new Zadko Observatory, which originally housed just the Zadko Telescope, was designed to incorporate up to six other telescopes situated on two metre high piers equally spaced around the viewing room. This was taken with the view that to ensure the future viability of the Observatory non-government sources of funding would be required.
We were fortunate through our relationship with the European Space Agency (ESA)  to host two ground based optical stations for Ariane used for space surveillance and space traffic management (\url{https://www.ariane.group}).

A USA based company Numerica Corporation have installed a ground based optical telescope for space debris and satellite tracking, plus general surveillance and observations for commercial and research purposes (\url{https://www.numerica.us/}).
As a part of a European Space Agency initiative ESA have requested time on the Zadko Telescope for follow-up observations of hazardous NEOs selected by the ESA NEOCC centre, with the most critical operations being NEO-alerts. This will require obtaining observations within 4-6 hours of a newly discovered NEO which has a non-zero probabilty of impacting the Earth (\url{https://www.esa.int/}).

Two piers house SPIRIT 3 and 4 telescopes, an initiative that incorporates a full life cycle of teacher and student professional learning opportunities and activities delivered via the outreach program at the International Centre for Radio Astronomy Research (ICRAR). SPIRIT 3 is a 35cm Schmidt-Cassegrain telescope manufactured by Celestron Telescopes. Its imaging camera is an Apogee Alta U6, which provides a 20 arc minute square field of view, and a default resolution of 1.2 arc seconds per pixel. SPIRIT 4 is a 32cm Corrected Dall-Kirkham telescope manufactured by Planewave Instruments. Its imaging camera is an SBIG STX-16803, offering a 50 arc minute field of view and a native resolution of 0.7 arc seconds per pixel.
ICRAR have also installed a telescope (Starfox) making use of next generation anti-reflection lens coatings to design complex new astrographs to minimise scattered light in the optical train optimising the instrument for low surface brightness imaging of the universe (\url{https://www.icrar.org/}).

\section{Conclusion}
After several years of upgrades the future is looking positive for the Zadko Observatory. Its unique location on the planet is becoming recognized internationally for time-domain astronomy. This is especially the case for gamma ray burst studies, space situational awareness and near Earth object follow-up. The Observatory now has a solid foundation to continue providing commercial services to international space agencies and companies.
Future projects of the Zadko Telescope include a complete upgrade to the Telescope Control System, installation of a new Guide Aquired Module allowing expanded capability for spectroscopy and the purchase of a new CMOS camera allowing for time domain photometry. These upgrades are expected to attract further commercial interest in the observatory.

\begin{figure}[!t]\centering
  \includegraphics[scale = 0.25,angle=-90]{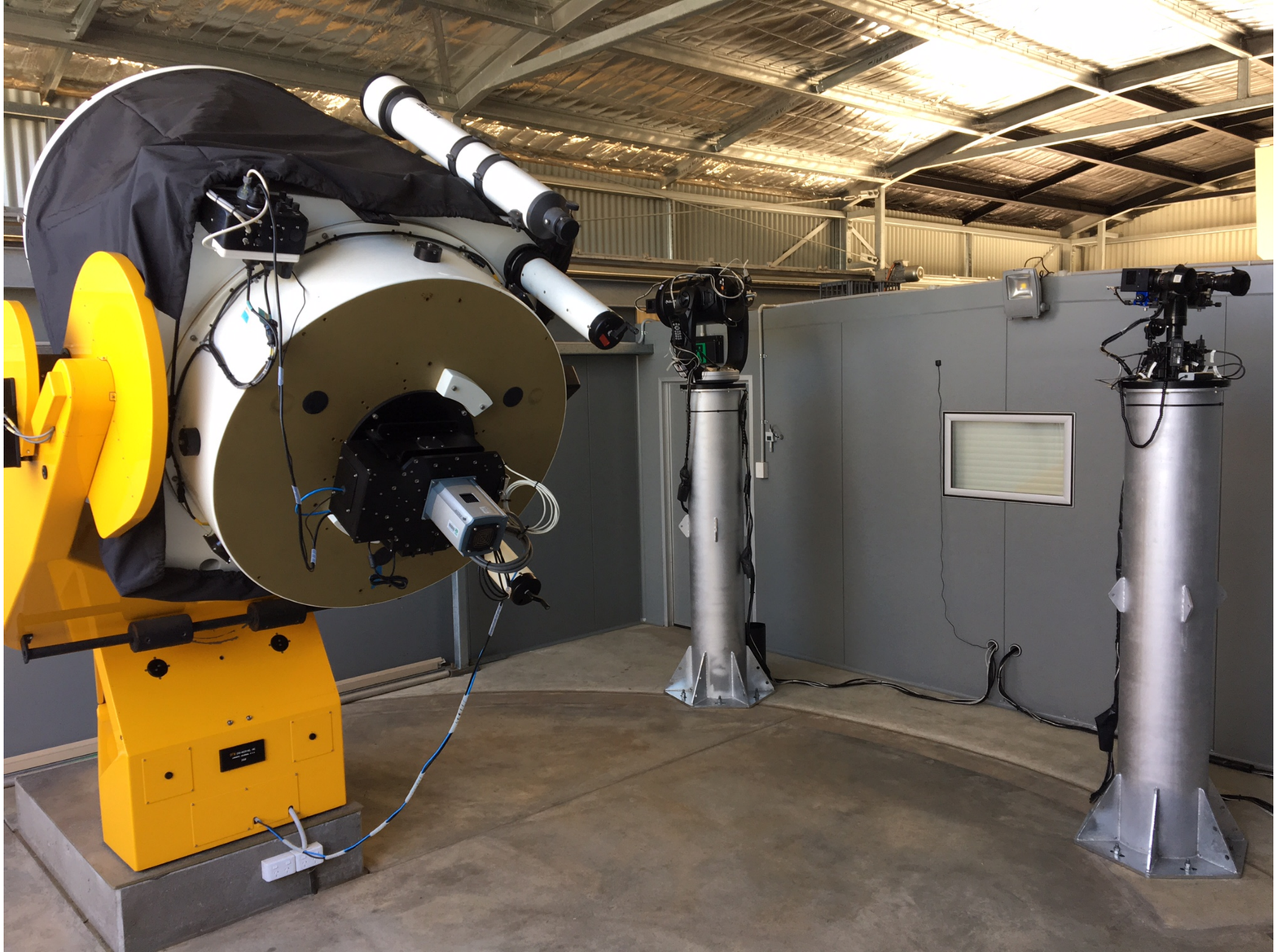}
  \caption{Zadko Telescope (foreground) along with several independent commercial instruments hosted inside the observatory.}
  \label{fig:telescope}
\end{figure}

\section{Acknowledgments}
The Zako Telescope receives partial support from the Australian Research Council Centre of Excellence for Gravitational Wave Discovery (OzGrav), project number CE170100004. We thank the Zadko family for providing funding for the Zadko Telescope and recognise the untimely passing of Jim Zadko.


\begin{thebibliography}
\bibitem[Abbott et al.(2017)]{abb17} Abbott, B. P., et al. 2017, \prl, 119, 161101
\bibitem[Andreoni et al.(2017)]{and17} Andreoni, I., et al. 2017, PASA, 34, 69
\bibitem[Antier et al.(2020)]{ant20} Antier, S., Agayeva, S., Aivazyan, V. et al. 2020, MNRAS, in press
\bibitem[Coward et al.(2010)]{cow10} Coward D., et al. 2010, PASA, 27, 331
\bibitem[Coward et al.(2017)]{cow17} Coward, D., et al. 2017, PASA, 34, 5
\bibitem[Gendre et al.(2020)]{gen20} Gendre B., Coward, D., Moore J. A., et al. 2020, RMAA, this volume
\end{thebibliography}
\end{document}